\documentclass[
twocolumn,
superscriptaddress,
amsmath,amssymb,
prl,
showkeys,
]{revtex4-2}
\usepackage{graphicx}
\usepackage{hyperref}
\usepackage{dcolumn}
\usepackage{bm}
\usepackage{epstopdf}
\usepackage{romannum}
\usepackage{csquotes}
\usepackage[dvipsnames]{xcolor}
\usepackage{makecell}
\DeclareUnicodeCharacter{2212}{\ensuremath{-}}
\usepackage{hyperref}
\usepackage{float}
\usepackage{graphicx} 

\usepackage{epsfig}
\usepackage{epstopdf} 

\begin{document}

\preprint{Sun/Anisotropy of FeTeSe}

\title{Impact of Disorder on the Superconducting Properties and BCS-BEC Crossover in FeSe Single Crystals}

\author{Jianping Fu}
\affiliation{Key Laboratory of Quantum Materials and Devices of Ministry of Education, School of Physics, Southeast University, Nanjing 211189, China}
\author{Yue Sun}
\email{Corresponding author: sunyue@seu.edu.cn}
\affiliation{Key Laboratory of Quantum Materials and Devices of Ministry of Education,	School of Physics, Southeast University, Nanjing 211189, China}
\author{Jingting Chen}
\affiliation{Department of Applied Physics, The University of Tokyo, Hongo, Bunkyo-ku, Tokyo 113-8656, Japan}
\author{Zhixiang Shi}
\email{Corresponding author: zxshi@seu.edu.cn}
\affiliation{Key Laboratory of Quantum Materials and Devices of Ministry of Education,	School of Physics, Southeast University, Nanjing 211189, China}
\author{Tsuyoshi Tamegai}
\affiliation{Department of Applied Physics, The University of Tokyo, Hongo, Bunkyo-ku, Tokyo 113-8656, Japan}

\date{\today}

\begin{abstract}
 
We investigate the crystal structure, transport properties and specific heat in five selected FeSe single crystals containing different amounts of disorder. Transport measurements show that disorder significantly suppresses superconducting transition temperature, $T_\mathrm{c}$, and upper critical field, $H_\mathrm{c2}$. Specific heat results confirm a robust multi-gap nature, a larger isotropic gap ($\Delta_\mathrm{s}$) and a smaller anisotropic gap ($\Delta_\mathrm{es}$). The smaller gap $\Delta_\mathrm{es}$ becomes more isotropic with increasing disorder. Additionally, FeSe is regarded as a superconductor in the crossover regime from Bardeen-Cooper-Schrieffer (BCS) to Bose-Einstein condensation (BEC) because of its comparable $\Delta$ and Fermi energy $E_\mathrm{F}$. By introducing disorder, the BCS-BEC crossover in FeSe can be tuned closer to BCS limit, reducing $\Delta/E_\mathrm{F}$ from 1.3 to 0.4. 

\end{abstract}

\maketitle

Traditional superconductivity is usually explained by the BCS theory, where weak electron-phonon interactions are the driving mechanism \cite{1964AP}. In these systems, the ratio of the superconducting transition temperature ($T_\mathrm{c}$) to the Fermi temperature ($T_\mathrm{F}$) is typically very small, on the order of $10^{-4}$ to $10^{-5}$. Even in high-$T_\mathrm{c}$ cuprates, this ratio remains around $10^{-2}$ at optimal doping levels. In contrast, BEC theory describes a state where a large number of bosons occupy the lowest quantum state as the temperature approaches absolute zero. In the BEC limit, the $T_\mathrm{c}/T_\mathrm{F}$ ratio increases to about $10^{-1}$ \cite{1995AM,2021HZ,2021PJ,2022KM}. In the BCS-BEC crossover regime, the strength of the pairing interaction increases, which bridges the weak-coupling Cooper pairs of the BCS theory and the tight-coupling bosons of the BEC theory. In this regime, $\Delta$ becomes comparable to $E_\mathrm{F}$. Meanwhile, the coherence length $\xi$ becomes comparable to reciprocal of Fermi wave vector $1/k_\mathrm{F}$, representing that the size of the Cooper pairs and the average interparticle distance are of the some order,  i.e. $1/k_\mathrm{F}\xi \sim 1$ \cite{2005CQ,2003GM,2004RC,2021BG,2021NY}. 

The BCS-BEC crossover exists in many strongly correlated superconductors. In underdoped high-$T_\mathrm{c}$ cuprates, the pseudogap formation has been discussed in terms of the BCS–BEC crossover \cite{2005CQ}. In low doping regime of Li$_x$ZrNCl, the phase diagram shows a pseudogap phase and $T_\mathrm{c}/T_\mathrm{F}$ is consistent with the edge of the BCS-BEC crossover regime \cite{2021NY}. In doped triangular system $\kappa$-(BEDT-TTF)$_4$Hg$_{2.89}$Br$_8$, the superconductivity crosses over from BEC limit to BCS limit with increasing pressure, which is attributed to a non-Fermi liquid to Fermi liquid (NFL-FL) crossover \cite{2022SY}. In Fe-based superconductor FeSe$_x$Te$_{1-x}$, The ratio $\Delta/E_\mathrm{F}\approx0.5$ is much larger than found in any previously studied superconductor, in agreement with the predictions of the BCS–BEC crossover theory \cite{KasaharaShigeru2014}. Particularly, previous studies suggest that FeSe may lie within this crossover regime, where $T_\mathrm{c}/T_\mathrm{F} \sim 0.1 - 0.2$, and $\Delta/E_\mathrm{F} \sim 1$ \cite{KasaharaShigeru2014,2023MY}. Therefore, FeSe-based superconductors provides an unique platform for understanding superconductivity with varying interaction strengths. However, it remains uncertain how disorder in FeSe single crystals affects this BCS-BEC crossover, as prior research has shown that superconductivity is always sensitive to disorder.

To properly investigate the BCS-BEC crossover, high-quality single crystals are crucial because the superconducting properties are sensitive to impurities and defects \cite{2018SY}. Therefore, in this study, we focus on the relationship between disorder(indicated by residual resistivity ratio, RRR) and superconducting properties in FeSe by investigating the composition, crystal structure, transport properties and specific heat of five selected FeSe single crystals with varying levels of disorder. Our findings suggest that as disorder increases, $E_\mathrm{F}$ increase monotonically, indicating that FeSe could be tuned in the BCS-BEC crossover regime.

FeSe single crystals were synthesized using the chemical vapor transport method \cite{Grow}. Five different types of crystals, labeled S1-S5, were selected from distinct batches. Notably, high-quality crystals were grown in a horizontal furnace, whereas lower-quality crystals were grown in a furnace tilted at 10°. The sensitivity of FeSe single crystal growth to temperature gradients implies that tilting the furnace introduces variable gradients, thereby affecting the overall quality of the crystals.
 
\begin{figure*}\center
	\centering
	\includegraphics[width=17cm]{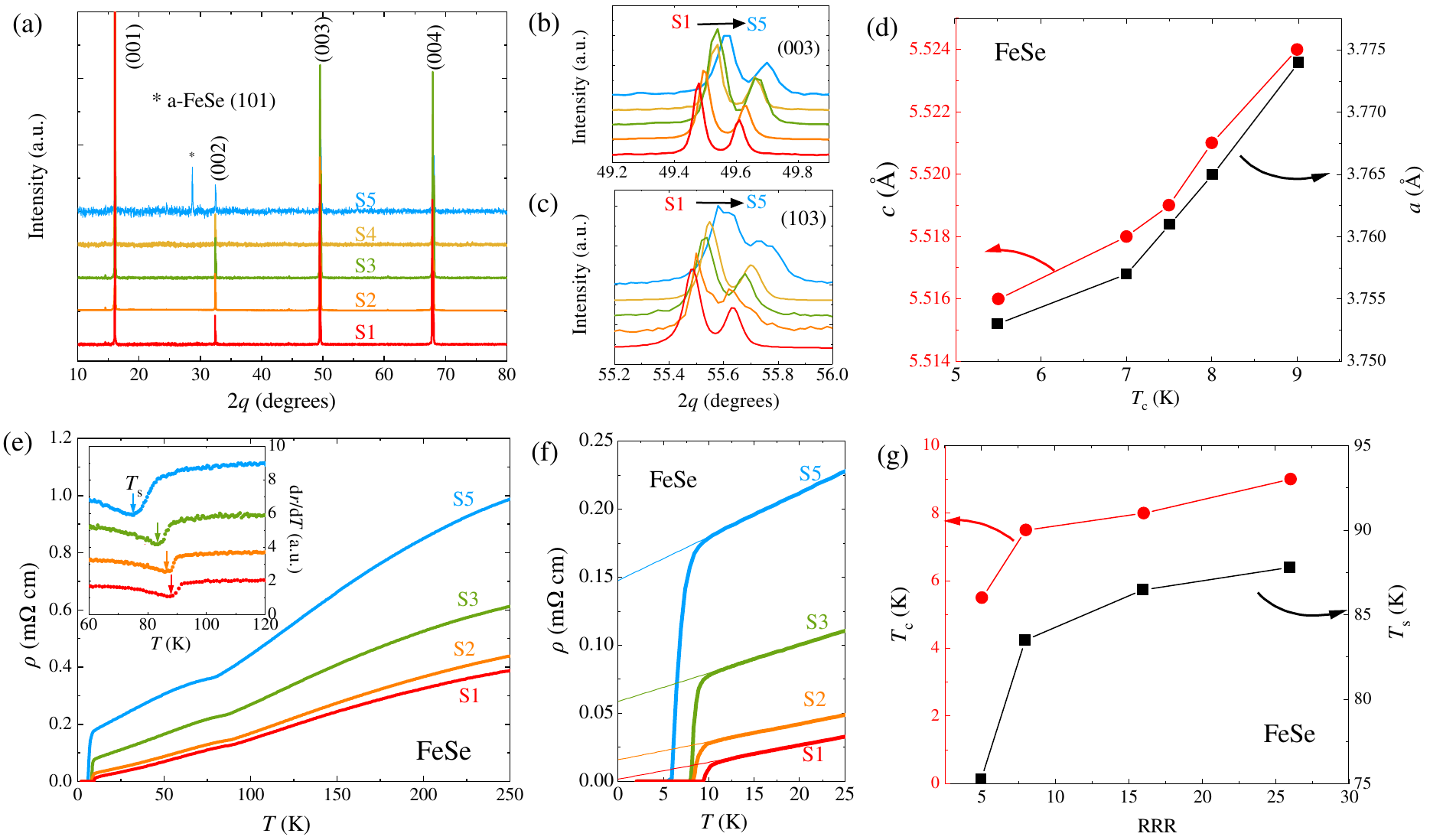}
	\caption{(a) XRD patterns for FeSe single crystals S1-S5, (003) and (103) peaks are enlarged in (b) and (c). (d) Evolution of the $a$ axis and $c$ axis for FeSe single crystals with different quality. (e) Temperature dependence of in-plane resistivity. (f) Larger view of (e) in the low-temperature region. (g) Plot of the $T_\mathrm{s}$ and $T_\mathrm{c}$ as a function of RRR for different FeSe crystals.}\label{fig1}
\end{figure*}

The structure of the synthesized crystals was characterized using X-ray diffraction (XRD) with Cu-K$\alpha$ radiation. The chemical compositions were analyzed using a scanning electron microscope (SEM) equipped with energy-dispersive X-ray spectroscopy (EDX). Magnetization measurements were conducted with a commercial superconducting quantum interference device (SQUID) magnetometer. Transport measurements were performed using the six-lead method with the applied field parallel to the $c$ axis and perpendicular to the applied current in a physical property measurement system (PPMS). Temperature dependence of specific heat was measured using thermal-relaxation method by an option of PPMS with $^3$He insert. This facility allows us to measure specific heat down to $\sim$ 0.5 K.

\begin{figure*}
	\centering
	\includegraphics[width=17cm]{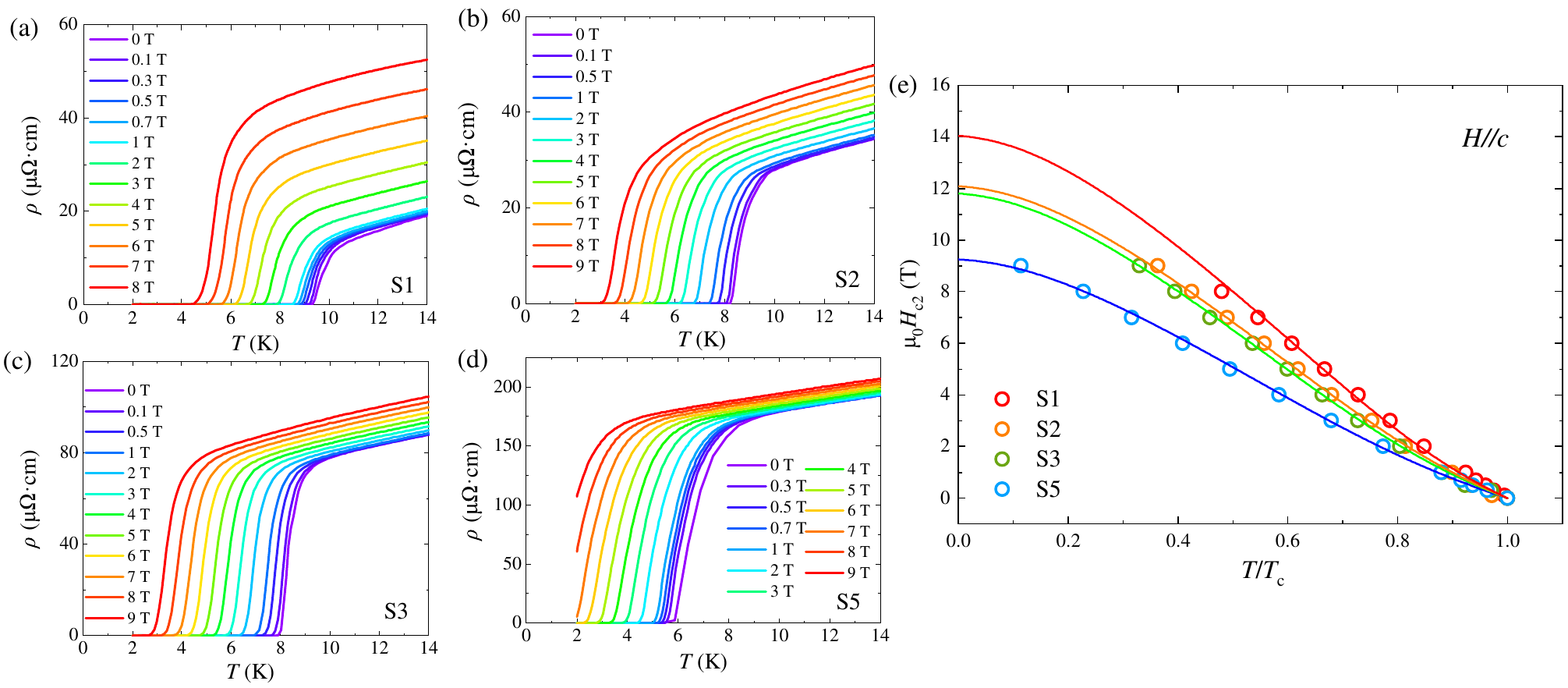}
	\caption{(a - d)Temperature dependence of resistivity in magnetic fields ($H \parallel c$) for S1, S2, S3, and S5. (e) Upper critical fields for FeSe single crystals as functions of normalized temperature. The solid line is two-band fitting for each crystal.}
	\label{fig2}
\end{figure*}

\begin{table*}
	\centering
	\caption{A summary of our experimental data of FeSe single crystals S1,2,3,5, including transport properties, lattice parameters along $a$ axis and $c$ axis, chemical composition, values of fitting coefficients $v_\mathrm{F}$ and $E_\mathrm{F}$ and two-band fitting data.}
	\begin{ruledtabular}
		\begin{tabular}{cccccccccccc}
			& RRR & $\rho_0$($\mu\Omega\cdot$cm) & $T_\mathrm{c}$(K) & $T_\mathrm{s}$(K) & $a$ axis($\AA$) & $c$ axis($\AA$) & Fe : Se & $v_\mathrm{F}$(m$\cdot^{-1}$) & $E_\mathrm{F}$(meV) & $\Delta$(meV) & $\xi$(nm) \\
			\colrule
			S1 & 26 & 2 & 9 & 87 & 3.774 $\pm$ 0.002 & 5.524 $\pm$ 0.002 & 0.995 : 1 & 1.07 $\times$ 10$^5$ & 2.29 & \makecell{ 1.69(71\%) \\ 0.33(29\%) \\ $\alpha$ = 0.72 } & 4.83      \\
			\hline
			S2 & 16 & 21 & 8 & 86 & 3.765 $\pm$ 0.001 & 5.521 $\pm$ 0.001 & 0.931 : 1 & 1.05 $\times$ 10$^5$ & 2.92 & \makecell{ 1.56(70\%) \\ 0.54(30\%) \\ $\alpha$ = 0.62 } & 5.20      \\
			\hline
			S3 & 8 & 60 & 7.5 & 84 & 3.761 $\pm$ 0.002 & 5.519 $\pm$ 0.002 & 0.920 : 1 & 1.05 $\times$ 10$^5$ & 4.35 & \makecell{ 1.36(65\%) \\ 0.50(35\%) \\ $\alpha$ = 0.60 } & 5.27      \\
			\hline
			S5 & 5 & 142 & 5.5 & 79 & 3.753 $\pm$ 0.002 & 5.516 $\pm$ 0.002 & 0.845 : 1 & 1.10 $\times$ 10$^5$ & 5.14 & \makecell{ 0.89(62\%) \\ 0.34(38\%) \\ $\alpha$ = 0.55 } & 5.95 \\
		\end{tabular}
	\end{ruledtabular}
\end{table*}\label{table.1}

Figure \ref{fig1}(a) displays the single crystal XRD patterns for the five FeSe crystals. The diffraction patterns show the ($00l$) peaks only, except for the impurity peak observed in S5, which is attributed to the non-superconducting hexagonal phase, $\alpha$-FeSe. This observation indicates that the crystallographic $c$ axis is perfectly perpendicular to the plane of the single crystals. Importantly, with increasing disorder, the positions of peaks shift to larger angle, which is more clearly illustrated by the enlarged (003) peaks in Fig. \ref{fig1}(b). This shift indicates a slight reduction in the $c$-axis lattice parameter.

To get deeper insight into the structural changes correlated with $T_{c}$, the diffraction angle of the (103) peak was specifically evaluated to calculate the $a$-axis lattice parameter, as depicted in Fig. \ref{fig1}(c). By using the following equations: 
\label{eq.1}
\begin{equation}
	d=n\lambda/2\mathrm{sin}\theta,
\end{equation}
and
\label{eq.2}
\begin{equation}
	d_{[hkl]}=\dfrac{1}{\sqrt{(\dfrac{h}{a})^2+(\dfrac{k}{b})^2+(\dfrac{l}{c})^2}}.
\end{equation}
The lattice parameters of the $a$ axis and $c$ axis for these crystals were calculated, as drown in Fig. \ref{fig1}(d). It is evident that both the $a$ axis and $c$ axis reduce with decreasing $T_{c}$, indicating shrinkage of the lattice. These observations are consistent with findings from other group \cite{FSS}. The EDX results presented in Table I indicate that the molar ratio of Fe in the four crystals is less than 1, suggesting that the amount of Fe is slightly lower than Se \cite{2018SY}. This lattice shrinkage, combined with a reduced amount of Fe, implies the presence of Fe vacancies in the FeSe single crystals. This phenomenon has also been confirmed through scanning tunneling microscopy (STM) measurements \cite{LinSTM}.

Fig. \ref{fig1}(e) illustrates the temperature dependence of in-plane resistivity for the FeSe single crystals with varying degrees of disorder, exhibiting metallic behavior until the temperature drops to $T_\mathrm{c}$. An enlarged view of the low-temperature $\rho - T$ curve is shown in Fig. \ref{fig1}(f). By linearly exploring the $\rho - T$ curve to 0 K, the residual resistivity $\rho_0$ can be determined, which directly correlates with the level of disorder. For crystal S1, $\rho_0$ is approximately 2 $\mu\Omega\cdot \text{cm}$, the smallest value among all reported values. Additionally, the residual resistivity ratio (RRR), defined as $\rho(250 \text{K})/\rho(10\text{K})$, is estimated to be approximately 26 for S1. This small residual resistivity $\rho_0$ and large RRR in S1 indicate minimal disorder. From S1 to S5, RRR decreases together with $T_\mathrm{c}$, suggesting disorder is harmful to superconductivity.

It is well recognized that FeSe single crystals undergo a structural transition from tetragonal to orthorhombic phases at the temperature $T_\mathrm{s}$, which is manifested as a kink-like behavior in the $\rho-T$ diagram. This structural transition is closely associated with $T_\mathrm{c}$ \cite{Grow,Structure}, where $T_\mathrm{s}$ exhibits a slight shift to lower temperatures as $T_\mathrm{c}$ decreases. The corresponding data for these four crystals are presented in Fig. \ref{fig1}(g). We observe that lower values of $T_\mathrm{s}$ and $T_\mathrm{c}$ correspond to a smaller RRR, closely similar to the findings of B{\"o}hmer et al. \cite{Grow}. In Fig. \ref{fig1}(g), RRR serves as a parameter of disorder and can be viewed as a 'tuning parameter'. As RRR decreases, both $T_\mathrm{s}$ and $T_\mathrm{c}$ also reduce as well. The smallest $\rho_0$ and the largest RRR indicate less disorder in S1, which aligns with its highest $T_\mathrm{c}$. Furthermore, in addition to the differences in $T_\mathrm{c}$, the specific heat at temperatures below 2 K varies among the four crystals, which will be discussed in detail later.

\begin{figure*}\center
	\includegraphics[width=17cm]{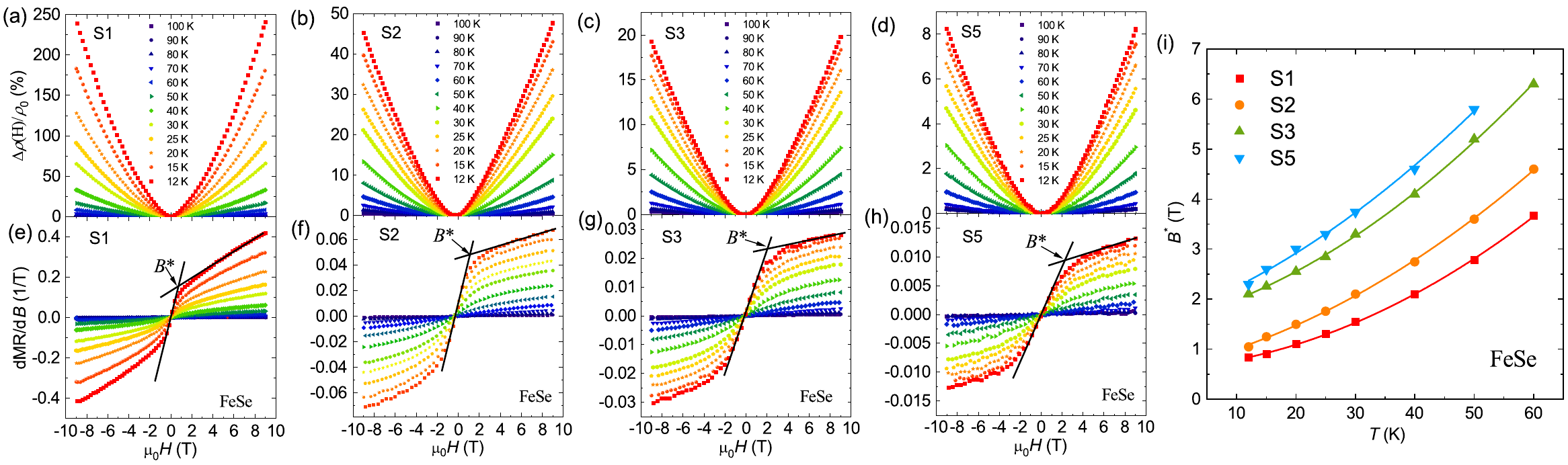}
	\caption{(a)-(d) Field dependence of MR and (e)-(h) Field derivative MR [d(MR)/dB] for S1, S2, S3, and S5, respectively. (i) Temperature dependence of the characteristic field $B^\ast$, and the solid lines represent the fitting for $B^\ast=(1/2e\hbar v_\mathrm{F}^2)(k_\mathrm{B} T+E_\mathrm{F})^2$.}\label{fig3}
\end{figure*}

The temperature dependence of resistivity under various magnetic fields for the four crystals S1, S2, S3, and S5 is illustrated in Figs. \ref{fig2}(a)-\ref{fig2}(d). The upper critical fields $H_\mathrm{c2}$ are defined as the temperature points where the resistivity drops to 50\% of the normal state value. Consequently, the temperature dependence of $\mu_0H_\mathrm{c2}$ is presented in Fig. \ref{fig2}(e). $\mu_0H_\mathrm{c2}(T)$ shows an almost linear temperature dependence and tends to be saturated at low temperatures, similar to the results of two-band superconductor MgB$_2$ \cite{2002GAA} and some IBSs \cite{2014PA,2023BM}, which has been successfully explained by the two-band model. To quantitatively analyze $\mu_0H_\mathrm{c2}$ data, the two-band model is expressed as follows:
\begin{equation}
	\begin{aligned}
         0 = &a_0(ln t + U(h))(ln t + U(\eta h)) + \\
         &a_1(ln t + U(h)) + a_2(ln t + U(\eta h)),
	\end{aligned}
\end{equation}
where $t=T/T_c$, $a_0=2(\lambda_{11}\lambda_{22}-\lambda_{12}\lambda_{21})/\lambda_{0}$, $a_1=1+(\lambda_{11}-\lambda_{22})/\lambda_{0}$, $a_2=1-(\lambda_{11}-\lambda_{22})/(\lambda_{0}/2)$, $\lambda_{0}=((\lambda_{11}-\lambda_{22})^2+4\lambda_{12}\lambda_{21})^{1/2}$, $h = \mu_0H_\mathrm{c2}D_1/(2\varPhi_0/T)$, $\eta=D_2/D_1$, $U(x)=\varPsi(1/2+x)-\varPsi(1/2)$. $\varPsi$ is digamma function, $D_1$ and $D_2$ are diffusivity of each band, $\lambda_{11}$, $\lambda_{22}$ denote the intra-band coupling constants, and $\lambda_{12}$, $\lambda_{21}$ are the inter-band coupling constants\cite{2017XXZ}. The values of $H_\mathrm{c2}$ at 0 K ($H_\mathrm{c2}$(0)) are 14.04 T, 12.09 T, 11.81 T, 9.25 T, for crystal S1, S2, S3, S5 respectively. Clearly, $H_\mathrm{c2}$(0) is gradually suppressed with the increase of disorder. Additionally, the coherence length $\xi$ determined using Ginzburg-Landau (GL) theory, $\mu_0H_\mathrm{c2}(T)=[\Phi_0/(2\pi\xi^2)][1-(T/T_c)]$,  is increased with disorder as presented in Table I.

To get deeper insight into the effect of disorder on the transport properties, the magnetoresistance (MR), defined as MR $= [\rho(H) - \rho(0)]/\rho(0)$, is measured, and shown in Figs. \ref{fig3}(a)-\ref{fig3}(d). Notably, the MR value in the crystal with $T_\mathrm{c}$ $\sim$ 9 K is very large, reaching nearly 250\% at 12 K under 9 T, indicating high quality. Meanwhile, MR value monotonically decreases in the four crystals, as disorder increases. 

In the semi-classical model, MR increases proportionally to $B^2$ in the entire field range. However, the MR in FeSe exhibits different trends, from a parabolic field dependence of MR at low fields to a more linear relation at high fields \cite{RN10}. This transition is evident in the first-order derivative d(MR)/d$B$, as shown in Figs. \ref{fig3}(e)-\ref{fig3}(h). In all four crystals, d(MR)/d$B$ linearly increases at low fields, while a reduced slope appears at high fields, attributed to a linear field-dependence of MR combined with a quadratic part \cite{KHH2014}. The intersection between these two behaviors occurs at a characteristic field $B^\ast$. One possible explanation for the reduced slope of d(MR)/d$B$ above $B^\ast$ is the competition in mobility between carriers from the normal band and those from the Dirac cone state within the multiband theory \cite{FSS,SY2014,2024QH}. Similar behavior is observed in BaFe$_2$As$_2$ and FeTe$_{0.6}$Se$_{0.4}$ \cite{SI2011,Chong_2013,RN4,RN10}. Notably, with the decrease in $T_\mathrm{c}$, the magnitude of MR tends to reduce, likely due to an decrease in carrier mobility, which is consistent with the increase in residual resistivity. 

The characteristic field $B^*$ defined as the intersect field between the semi-classical regime and the linear regime, is marked by the arrows in Figs. \ref{fig3}(e)-\ref{fig3}(h). The temperature dependence of $B^*$ of four selected crystals is illustrated in Fig. \ref{fig3}(i). This relationship is obviously violating the linear relation expected from conventional parabolic bands, and can be expressed as  $B^\ast=(1/2e\hbar v_\mathrm{F}^2)(k_B T+E_\mathrm{F})^2$ for Dirac fermions, where $v_\mathrm{F}$ represents the Fermi velocity and $E_\mathrm{F}$ denotes the Fermi energy, depicted by the solid line. The good agreement of $B^*$ with the above equation confirms the existence of Dirac fermions in FeSe. Notably, there is only a minimal variation in $v_\mathrm{F}$ across these four crystals, suggesting the stability of the Dirac-cone-like band structure in FeSe. It is reasonable to say that the band structure changes little by disorder. The magnitude of $E_\mathrm{F}$ in crystal S1 is consistent with the values reported by the quantum oscillations and ARPES measurements \cite{2014TT,KasaharaShigeru2014}. However, a notable increase in Fermi energy from 2.29 to 5.14 meV is observed with decreasing $T_\mathrm{c}$.

\begin{figure*}\center
	\includegraphics[width=17cm]{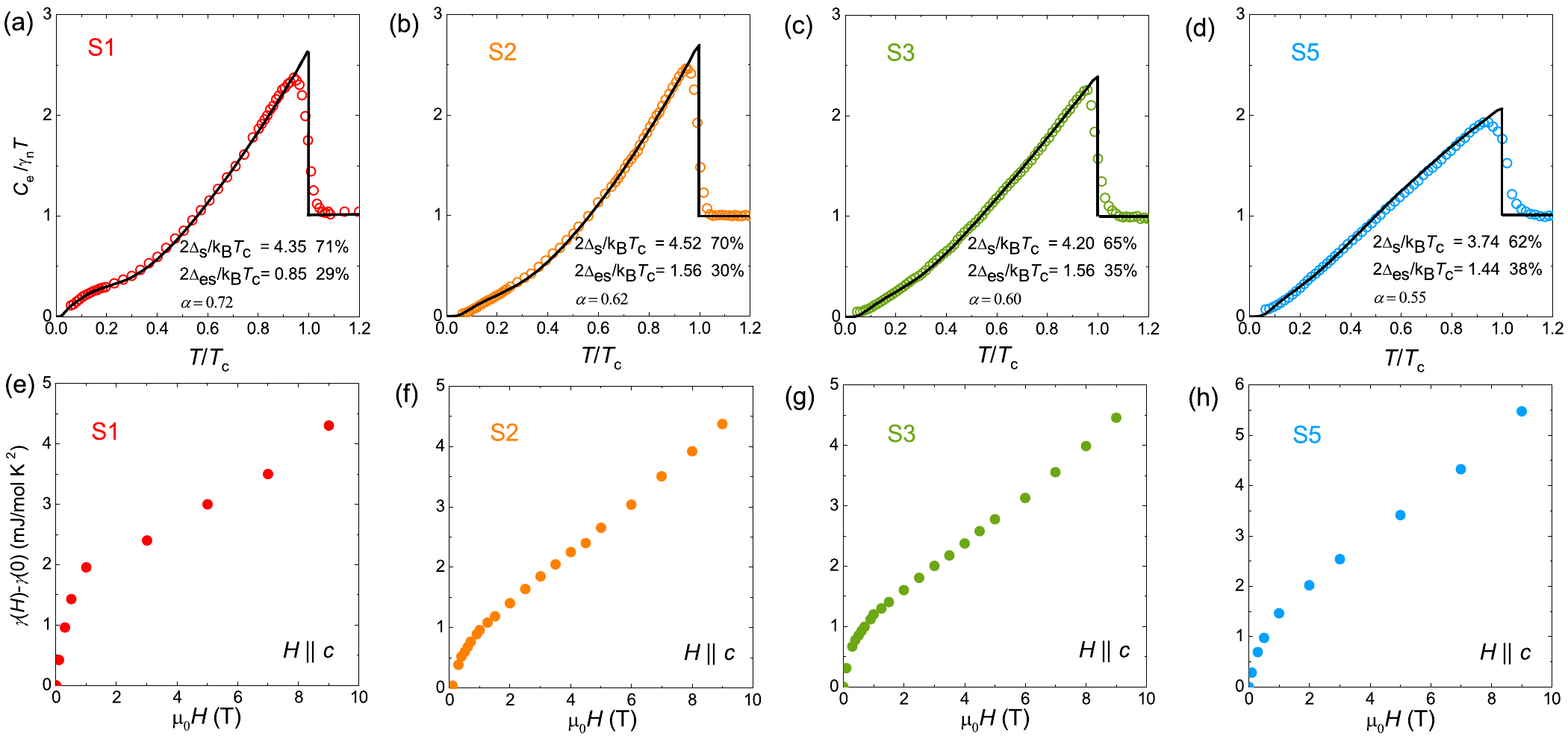}
		\caption{(a)-(d) $s$ + extended $s$-wave fittings for the temperature of electronic specific heat of FeSe single crystals S1, S2, S3, and S5. (e)-(h) Magnetic field dependence of electronic specific heat coefficient in four crystals.}\label{fig4}
\end{figure*}

Zero-field electronic specific heat $C_\mathrm{e} /T$ is obtained by subtracting the phonon terms under 9 T (details in Supplemental Information), shown in Fig. \ref{fig4}. A clear jump associated with the superconducting transition is observed, which is consistent with the resistivity measurement in Fig. \ref{fig2}. Generally, the specific heat jump reduces with increasing disorder, similar to other superconductors \cite{GH1997}. From Fig. \ref{fig4}(a), other than the specific heat jump at $T_\mathrm{c}$ $\sim$ 9 K, a hump-like behavior is also observed at $\sim$ 1.2 K in S1. This is a typical behavior of a two-gap superconductor such as MgB$_2$ \cite{FB2001} and Lu$_2$Fe$_3$Si$_5$ \cite{2012YN}. Such two-gap nature, as well as a disorder-sensitive small gap have been well studied by angle-resolved specific heat (ARSH) measurement in our previous report \cite{2017SY}. Besides, a very weak second drop is also discovered in S2 while the similar second-drop feature is strongly suppressed in other two crystals, S3 and S5, as shown in Figs. \ref{fig4}(b)-\ref{fig4}(d). 

In order to get deeper understanding of the gap structure in FeSe, it is necessary to fit the data of electronic specific heat with two-gap $\alpha$-model:
\label{eq.3}
\begin{equation}
\begin{aligned}
	C_\mathrm{es}=&\frac{4N(0)}{k_BT^2}\int_0 ^{\hbar{h} \omega_D}d\varepsilon[\varepsilon^2+\Delta^2-\frac{T}{2}\frac{d\Delta^2}{dT}]\\
	&(1+e^{E/k_BT})^{-2}e^{E/k_BT},
\end{aligned}
\end{equation}
where $N(0)$ is the density of states at the Fermi level, $E = (\varepsilon^2 + \Delta^2)^{1/2}$, $\Delta$ is the superconducting energy gap. For isotropy $s$-wave, $\Delta_\mathrm{s}=\Delta_0$. For extended $s$-wave, the order parameter used to fit the data is $\Delta_\mathrm{es} = \Delta_0 (1+\alpha \mathrm{cos}2\theta)$ ($\alpha$ donates the gap anisotropy). Apparently, two-gap fitting leads to satisfactory results as shown in Figs. \ref{fig4}(a)-\ref{fig4}(d), indicating two-gap nature in the four crystals. Notably, the ratio $2\Delta_\mathrm{s}/k_\mathrm{B}T_\mathrm{c}$ for the larger band is always lager than BCS limit 3.53, showing strong coupling superconductivity. Superconducting gap parameters and contribution from each gap are summarized in Table I. The coefficient $\alpha$ slightly reduces, indicating that the extended $s$-wave becomes more isotropic with increasing disorder. This is because the disorder scattering can lift the gap minimum \cite{2012JR}. The coexistence of a small anisotropic $s$-wave gap and a large more isotropic $s$-wave gap in FeSe single crystals have also been confirmed by angle-resolved photoemission spectroscopy (ARPES) measurements \cite{KasaharaShigeru2014,2017RS,2017SY}.  

More information about the gap structure can be obtained from the magnetic field dependence of the specific heat (see supplementary information), which reflects the quasiparticle excitation. By extrapolating Fig. S3 and Fig. S4 to $T$ = 0 K, the field dependence of $\gamma(H)-\gamma(0)$ are obtained and shown in Figs. \ref{fig4}(e)-\ref{fig4}(h). The clear steep increase in $\gamma(H)$ in S1 at low fields indicates the quick suppression of smaller gap by magnetic fields, while such behavior is greatly suppressed in other three crystals. For Figs. \ref{fig4}(a)-\ref{fig4}(d), $\Delta_\mathrm{es}$ in S1 is extremely small and highly anisotropic, indicating the quasiparticles can be easily excited, while $\Delta_\mathrm{es}$ in other three crystals are nearly two times larger and more isotropic. Field dependence of $C/T$ further confirms the multi-gap nature. 

\begin{figure}\center
	\includegraphics[width=8.5cm]{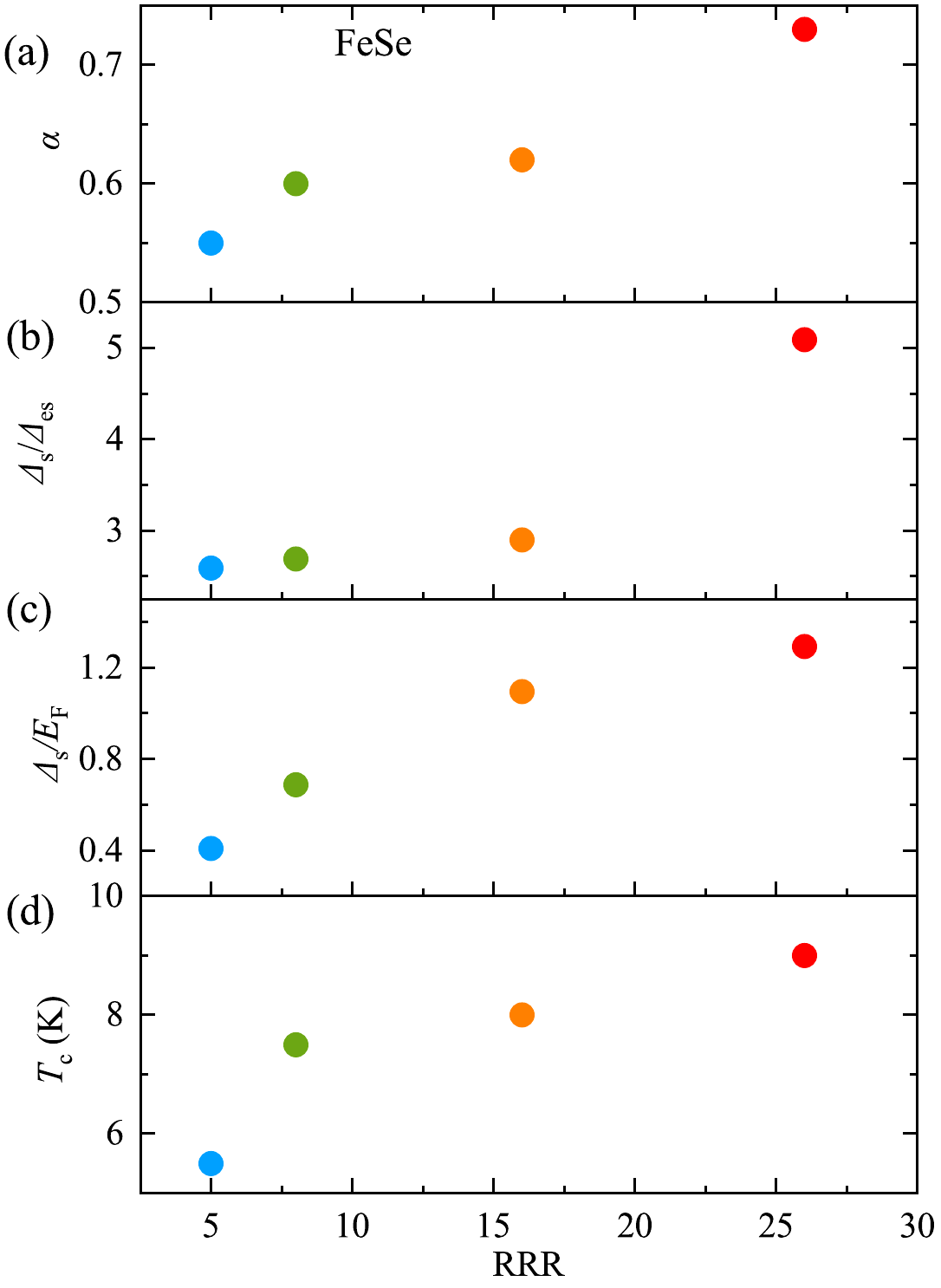}
	\caption{RRR dependence of (a) gap anisotropy $\alpha$ extracted from Fig. \ref{fig4}, (b) ratio of two gaps, (c) the ratio of $\Delta_\mathrm{s}/E_\mathrm{F}$, and (d) superconducting transition temperature.}\label{fig5}
\end{figure}

Fig. \ref{fig5} depicts the evolution of gap anisotropy $\alpha$, $\Delta_\mathrm{s}/\Delta_\mathrm{es}$ and $\Delta_\mathrm{s}/E_\mathrm{F}$, as a function of RRR. Generally, $\Delta_\mathrm{s}/\Delta_\mathrm{es}$ becomes smaller with the decrease in RRR, suggesting a pronounced correlation between gap anisotropy and RRR. In addition, $\Delta/E_\mathrm{F}$ can be converted to the dimensionless crossover parameter ln($k_\mathrm{F}a_\mathrm{2D}$), where $a_{2D}$ is the 2D scattering length \cite{2015HL,2000CQ,2006RS,2006PG}. The edge of crossover regime in 2D systems corresponds to ln($k_\mathrm{F}a_\mathrm{2D}$) $>$ 2 and $\Delta/E_\mathrm{F}$ $>$ 0.3, while the conventional BCS theory relies on the relation $k_\mathrm{B} T_\mathrm{c} \sim \Delta \ll E_\mathrm{F}$. The monotonic increase in $\Delta/E_\mathrm{F}$ toward the low disorder limit represents the continuous shift from the BCS limit to a more BEC limit. Note that $\Delta/E_\mathrm{F}$ of four crystals S1, S2, S3, S5 are all larger than 0.3, corresponding to the BCS-BEC crossover region \cite{2015HL,2021NY}. Consequently, the system undergoes a transition from a Fermi liquid to a Bose liquid, signifying a BCS-BEC crossover.

Now we discuss the issue of the BCS-BEC crossover. The physical description of the zero-temperature BCS-BEC crossover is primarily based on two key quantities: the condensate fraction \cite{2005SL,2020HT,2020AD} and the average size of Cooper pairs \cite{1994PF,2018SG}. The condensate fraction represents the proportion of fermions that participating in the condensation process, forming the superconducting state. In the BCS regime, this fraction is exponentially small, meaning only a minimal density of fermions contributes to the condensate wavefunction. In contrast, in the BEC regime, the condensate fraction approaches unity, where the attraction between fermions is strong enough to form tightly bound molecular pairs with bosonic character. The average pair size, $\xi$, clarifies the distinction between the two regimes. The parameter $1/k_\mathrm{F}\xi$, introduced in pioneering studies on the BCS-BEC crossover \cite{1994PF,2014AG}, serves as a crucial indicator. In the BCS regime, $1/k_\mathrm{F}\xi \ll 1$, signifying large Cooper pairs, whereas in the BEC regime, $1/k_\mathrm{F}\xi > 1$, reflecting point-like pairs. This framework provides a clearly theoretical and phenomenological understanding of the BCS-BEC crossover, which is consistent to previous report \cite{KasaharaShigeru2014}.

In FeSe crystal, $k_\mathrm{F}$ obtained from reference \cite{2014TT}, is roughly 0.43 nm$^{-1}$.  $\xi$ is determined from the upper critical field, as shown in Table I. Since $v_\mathrm{F}$ keeps almost unchanged with increasing disorder, we assume that $k_\mathrm{F}$ is almost constant. With disorder compensated, $1/k_\mathrm{F}\xi$ monotonically increase from 0.39 to 0.48. The correlation parameter $1/k_\mathrm{F}\xi$ is $\sim$ 0.48 meaning a few Cooper pairs overlap, contrary to the assumption in the BCS theory \cite{2004ZM}. We assume that disorder may increase electronic interaction, driving FeSe to BEC limit. Therefore, it is worth exploring the potential manifestations of the BCS-BEC crossover in FeSe, although these manifestations might significantly differ from those expected in single-band superconductors.

Our experimental results demonstrate that disorder significantly affects the superconducting properties of FeSe single crystals. Specifically, with increasing disorder, both the $a$-axis and $c$-axis lattice parameters decrease, residual resistivity increases, and $T_\mathrm{c}$ significantly decreases. Disorder introduces local stress and electron scattering centers, affecting the electronic structure and coherence length in FeSe, thereby suppressing superconductivity. Additionally, our study supports that FeSe resides in the BCS-BEC crossover regime. In this regime, superconducting pairing transitions from the weakly coupled BCS state to the strongly coupled BEC state, with FeSe exhibiting high sensitivity to disorder. In summary, our research not only validates previous findings but also provides additional experimental data on the impact of disorder on the superconducting properties of FeSe, offering important insights into the superconducting mechanism of FeSe. Moreover, our study supports that FeSe is situated in the BCS-BEC crossover regime, providing experimental evidence for exploring novel physical phenomena within this crossover.

\acknowledgements
This work was partly supported by the National Key R$\&$D Program of China (Grants No. 2024YFA1408400) and the National Natural Science Foundation of China (Grants No. 12374136, No. 12374135, No. U24A2068, No. 12204487).

J.F. and Y.S. contributed equally to this work.


\bibliography{references}

\end{document}